\documentclass[conference]{IEEEtran}
\IEEEoverridecommandlockouts

\usepackage{cite}
\usepackage{graphicx}
\usepackage{amsmath}
\usepackage{array}
\usepackage{mdwmath}
\usepackage{amssymb}
\usepackage{mdwtab}
\usepackage{stfloats}
\usepackage[tight,footnotesize]{subfigure}
\usepackage{amsmath,amsthm}
\usepackage{threeparttable}
\usepackage{color}
\usepackage{url}
\usepackage{algpseudocode}
\usepackage{algorithm}
\usepackage{multicol}
\usepackage{amssymb}
\usepackage{epstopdf}
\usepackage{flushend}
\usepackage[letterpaper, right=0.62in, left = 0.63in, top = 1.92cm, bottom = 1.7in]{geometry}

\algrenewcommand{\algorithmicrequire}{\textbf{Input:}}
\algrenewcommand{\algorithmicensure}{\textbf{Output:}}

\newtheorem{lemma}{Lemma}

\def\BibTeX{{\rm B\kern-.05em{\sc i\kern-.025em b}\kern-.08em
    T\kern-.1667em\lower.7ex\hbox{E}\kern-.125emX}}

\begin{document}
\title{Performance Evaluation of RIS-Assisted Spatial Modulation for Downlink Transmission}
\author{
	\IEEEauthorblockN{
       Xusheng Zhu, Qingqing Wu, Wen Chen
}
	\IEEEauthorblockA{Department of Electronic Engineering, Shanghai Jiao Tong University, Shanghai, China}
	\IEEEauthorblockA{Email: \{xushengzhu, qingqingwu, wenchen\}@sjtu.edu.cn}
}

\markboth{}
{}

\maketitle
\begin{abstract}
This paper explores the performance of reconfigurable intelligent surface (RIS) assisted spatial modulation (SM) downlink communication systems, focusing on the average bit error probability (ABEP). Notably, in scenarios with a large number of reflecting units, the composite channel can be approximated by a Gaussian distribution using the central limit theorem. The receiver utilizes a maximum likelihood detector to recover information in both spatial and symbol domains. In the proposed RIS-SM system, we analytically derive a closed-form expression for the union tight upper bound of ABEP, employing the Gaussian-Chebyshev quadrature method. The validity of these results is rigorously confirmed through exhaustive Monte Carlo simulations.
%
\end{abstract}

\section {Introduction}

In recent years, reconfigurable intelligent surface (RIS) has emerged as a pivotal technology in wireless communication networks, owing to its cost-effectiveness, passive nature, and ease of deployment \cite{bash2021reconf}. Specifically, RIS can enhance communication quality by precisely adjusting the phase and amplitude of each element to create reflective beams directed towards target objects \cite{wu2019inte}.
Spatial modulation (SM), on the other hand, is a promising modulation technology that maps information not only to phase shift keying/quadrature amplitude modulation (PSK/QAM) symbols but also to the index of the antenna \cite{zhu2022on}. By incorporating the antenna index into signal modulation, the system's spectral efficiency can be significantly improved \cite{jeg2003spatial}. Notably, the SSK scheme presented in \cite{jegan2009space,zhuicccda,ris2023zhu} emphasizes reliability by neglecting the symbol domain PSK/QAM from SM.

Given the advantages of both RIS and SM, there is growing interest in their combination and various variants \cite{basar2020rec,ma2020large,luo2021spatial,zhu2023rissk,zhu2023RIS,zhucsi2022,zhu2024fdsk}.  For instance, in \cite{basar2020rec}, RIS-SSK and RIS-SM schemes were proposed, deploying RISs close to the transmitter and implementing spatial domain modulation indices at the receive antenna. These schemes enhance average bit error probability (ABEP) performance at low signal-to-noise ratio (SNR) regions and increase spectral efficiency.
\cite{ma2020large} explored RIS-SM, deriving its ABEP performance and analytical expression using an approximating Q-function. \cite{luo2021spatial} investigated RIS-assisted SM in the uplink, formulating an optimization problem to reduce the symbol error rate. Moreover, \cite{zhu2023rissk} introduced the RIS-SSK scheme with the RIS deployed in the middle of the channel, and \cite{zhucsi2022} considered the impact of channel estimation errors on reliability. In the millimeter wave, \cite{zhu2023RIS,zhu2024mMIO} investigated the ABEP performance on RIS assisted spatial scattering modulation (SSM) schemes.

Against the above background,
we propose the RIS-assisted SM scheme for downlink communication transmission.
To the best of our knowledge, there is no existing literature on the RIS-SM scheme for downlink communication systems.
In contrast to  \cite{basar2020rec}, we consider a more prevalent RIS-SM model where the RIS is positioned in the middle of the channel, and the SM is implemented at the transmitter, as opposed to implementing SM at the receiver side.
Unlike \cite{ma2020large}, we study the two metrics, namely ABEP and ergodic capacity, separately, giving closed-form expressions for each.
Different from \cite{luo2021spatial}, our focus is on exploring the RIS-SM scheme for downlink transmission.
It is noteworthy that the signal model considered by  \cite{zhu2023rissk,zhu2024fdsk,zhucsi2022} do not account for PSK/QAM in the symbol domain, making it a special case of the proposed scheme.
In summary, the contributions of this paper can be described as follows:
1) In this paper, we investigate a RIS-assisted SM downlink transmission scheme. At the receiver side, we employ the maximum likelihood (ML) detector to recover the original information.
2) Based on ML detector, we first derive
the closed-form union upper bound of ABEP via Gaussian-Chebyshev quadrature (GCQ) method.
3) The simulation results are used to verify the correctness of the theoretical derivation.

\emph {Notations:}
Lowercase bold letters and uppercase bold letters denote vectors and matrices, respectively.
$(\cdot)^T$, $(\cdot)^H$, and $(\cdot)^*$ are the transposition and Hermitian transposition, respectively.
${\rm diag}(\cdot)$ denotes the diagonal matrix operation.
$\mathcal{N}(\cdot,\cdot)$ and $\mathcal{CN}(\cdot,\cdot)$ denote the real and complex Gaussian distributions, respectively.
$\Re$ and $\Im$ represent the real and imaginary part operations, respectively.
$E[\cdot]$ and $Var[\cdot]$ indicate the expectation and variance operations, respectively.

\begin{figure}
  \centering
  \includegraphics[width=6.0cm]{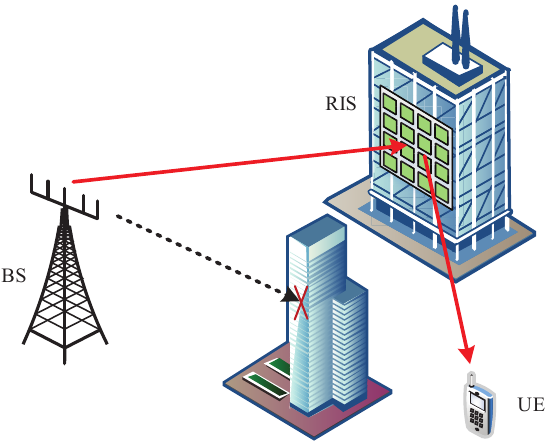}
  \caption{\small{The RIS-assisted SM system model.}}\label{sys}
\end{figure}
\section {System Model}
In this paper, we consider the RIS-assisted SM downlink transmission system as shown in Fig. \ref{sys}, which is composed of a base station (BS), a user equipment (UE), and a piece of RIS.
Due to the existing building blockage between the BS and UE, we assume that there is no direct link from the BS to UE.
For this reason, we resort to a RIS to assist the downlink signal transmission from the BS to the UE.
In Fig. \ref{sys}, we assume that the BS and UE are equipped with $N_t$ and a single antenna for reception, respectively.
Besides, the RIS consists of $L$ low-cost passive reflection units, each of which can adjust the amplitude and phase of the reflected signal independently.
To characterize the performance limit of RIS, we assume that the reflection amplitude per unit number is one.
In the following, we utilize the matrix $\boldsymbol{\Theta} ={\rm diag}\{e^{j\theta_1},e^{j\theta_2},\cdots,e^{j\theta_L}\}$ to represent RIS reflection coefficient, where $\theta_l, l\in\{1,2\cdots,L\}$ is the reflection phase shift of the $l$-th unit on the RIS.

The wireless channels between the UE and RIS, and RIS and $n_t$-th antenna of BS are respectively described as $\boldsymbol{g}\in\mathbb{C}^{L\times 1}$ and
$\boldsymbol{h}_{n_t}\in\mathbb{C}^{L\times 1}$, where $g_l =\beta_le^{-j\phi_l}$ and $h_{l,n_t} =\alpha_{l,n_t}e^{-j\varphi_{l,n_t}}$ denote the channels between the $l$-th reflecting element of RIS and UE and $l$-th reflecting element and BS, respectively.
Note that $\beta_l$ and $\phi_l$, $\alpha_{l,n_t}$ and ${\varphi_{l,n_t}}$ represent the amplitudes and phases of the fading channels $g_l$ and $h_{l,n_t}$, respectively.
To explore the achievable performance of SM, we assume that the BS and UE can obtain perfect channel state information (CSI). For the required reflection coefficien, the BS sends it to the RIS controller through a wired or wireless link. It is worth noting that the RIS controller can tune the reflection state of each RIS element in real time based on the information received from the BS side.
At each time slot, the input message $\log_2(N_t)+\log_2(M)$ consists of two parts. The first $\log_2(M)$ bit maps to a constellation point $s$ in the $M$-PSK/QAM signal selected from the symbol domain code book $\mathcal{S}= \{s_1,\cdots, s_m,\cdots,s_M\}$, where the corresponding average power with respect to $s$ satisfies $E[|s|^2]=1$.
The rest of the $\log_2(N_t)$ bits come from the spatial domain constellation point, which is selected by the spatial domain index set $\mathcal{N}=\{1,2,\cdots,{N_t}\}$.
In this way, the received signal at the UE side is given by
\begin{equation}\label{received}
\begin{aligned}
y &= \sqrt{P_t}\boldsymbol{g\Theta}\boldsymbol{h}_{n_t}s + n_0\\
&=\sqrt{P_t}\sum\nolimits_{l=1}^L\alpha_{l,n_t}\beta_le^{j(\theta_l-\phi_l-\varphi_{l,n_t})}s + n_0,
\end{aligned}
\end{equation}
where $P_t$ denotes the average transmit power and $n_0\sim\mathcal{CN}(0,N_0)$ represents the additive white Gaussian noise (AWGN).
To maximize the instantaneous SNR value, the transmission signal quality can be improved with the help of RIS. Specifically, by adjusting $\theta_l$ to combine $\phi_l$ and $\varphi_{l,n_t}$ the channel phase is zero ,i.e., $\theta_l=\phi_l+\varphi_{l,n_t}$.
In this respect, (\ref{received}) can be rewritten as
\begin{equation}
\begin{aligned}
y =\sqrt{P_t}\sum\nolimits_{l=1}^L\alpha_{l,n_t}\beta_ls + n_0.
\end{aligned}
\end{equation}
At the UE side, the ML detector jointly estimates the transmit antenna index $\hat n_t$ and the symbol information $s$ as
\begin{equation}\label{ml}
[\hat n_t, \hat s] = \arg\min\limits_{n_t\in\mathcal{N}, s\in\mathcal{S}}|y-\sqrt{P_t}\sum\nolimits_{l=1}^{L}\alpha_{l,n_t}\beta_l s|^2.
\end{equation}

\section {Performance Analysis}
In this section, the analytical expressions of ABEP are given over the Rayleigh fading channels based on the ML detector.
\subsection{Conditional Pair Error Probability (CPEP)}
To gain an upper bound on ABEP, we first derive the CPEP expression as
\begin{equation}\label{cpep1}
\begin{aligned}
P_b 
=&\Pr(|y-\sqrt{P_t}\sum\nolimits_{l=1}^{L}\alpha_{l,n_t}\beta_l s|^2\\&>|y-\sqrt{P_t}\sum\nolimits_{l=1}^L\alpha_{l,\hat n_t}\beta_{l}e^{-j(\theta_{l,\hat n_t}-\theta_{l,n_t})}\hat s|^2)\\
\overset{(a)}{=}&\Pr(|y-\sqrt{P_t}G_{n_t} s|^2>|y-\sqrt{P_t}G_{\hat n_t}\hat s|^2)\\
=&\Pr(\left|n_0\right|^2>\left|\sqrt{P_t}(G_{ n_t} s-G_{\hat n_t}\hat s\right)+n_0|^2)\\
=&\Pr(-\left|\sqrt{P_t}\left(G_{ n_t} s-G_{\hat n_t}\hat s\right)\right|^2\\&-2\Re\{n_0^H\sqrt{P_t}\left(G_{ n_t} s-G_{\hat n_t}\hat s\right)\}>0)\\
=&\Pr\left(F>0\right),
\end{aligned}
\end{equation}
where $(a)$ denotes the $G_{n_t}=\sum_{l=1}^{L}\alpha_{l,n_t}\beta_l$ and $G_{\hat n_t}=\sum_{l=1}^L\alpha_{l,\hat n_t}\beta_{l}e^{-j(\theta_{l,\hat n_t}-\theta_{l,n_t})}$.
Note that $F=-\left|\sqrt{P_t}\left(G_{ n_t} s-G_{\hat n_t}\hat s\right)\right|^2-2\Re\{n_0^H\sqrt{P_t}\left(G_{ n_t} s-G_{\hat n_t}\hat s\right)\}$ is a real Gaussian random variable.
Accordingly, the mean and variance values of $F$ can be given as
$
\mu_F=-\left|\sqrt{P_t}\left(G_{ n_t} s-G_{\hat n_t}\hat s\right)\right|^2$ and $
\sigma_F^2=2N_0P_t\left|G_{ n_t} s-G_{\hat n_t}\hat s\right|^2
$
Herein, the CPEP can be calculated as
\begin{equation}\label{cpep2}
P_b=Q(-\mu_F/\sigma_F^2)=Q\left(\sqrt{\frac{\rho\left|G_{ n_t} s-G_{\hat n_t}\hat s\right|^2}{2}}\right),
\end{equation}
where $\rho = {P_t}/{N_0}$ stands for the average SNR.
\subsection{Unconditional Pair Error Probability (UPEP)}
\subsubsection{Correct transmit antenna detection $\hat n_t = n_t$}
In this case, the (\ref{cpep2}) can be simplified as
\begin{equation}
P_b=Q\left(\sqrt{\frac{\rho|G_{ n_t}|^2| s-\hat s|^2}{2}}\right).
\end{equation}
For the sake of simplicity, we define $x=|G_{ n_t}|^2$.
In this way, the UPEP can be given as
\begin{equation}\label{cpep3}
\begin{aligned}
\bar P_b &= \int_0^\infty f(x)Q\left(\sqrt{\frac{\rho x|s-\hat s|^2}{2}}\right)dx,
\end{aligned}
\end{equation}
where $f(x)$ denotes the probability density function (PDF) with respect to the variable $x$. Since it is difficult to find the distribution of $x$ directly, we first calculate the distribution of $\xi = \sum_{l=1}^{L}\alpha_{l,n_t}\beta_l$. Note that the distribution of $f(x)$ is obtained on the basis of the distribution of $\xi\sim\mathcal{N}\left(\mu_\xi, \sigma_\xi^2\right)$,
where the mean and variance can be denoted as $\mu_\xi=\frac{L\pi}{4}$ and $\sigma_\xi^2=\frac{L(16-\pi^2)}{16}$, respectively.
It is worth noting that the moment generating function (MGF) of variable $x$ can be given as \cite{zhucsi2022}
\begin{equation}\label{xpdf}
\begin{aligned}
&{\rm MGF}_\Gamma(x)=\sqrt{\frac{1}{1-2x\sigma_\xi^2}}\exp\left(\frac{x\mu_\xi^2}{1-2x\sigma_\xi^2}\right),
\end{aligned}
\end{equation}
Taking (\ref{xpdf}) into (\ref{cpep3}), the UPEP can be evaluated as
\begin{equation}\label{epb1}
\begin{aligned}
\bar P_b
=&\frac{1}{\pi}\int_0^{\frac{\pi}{2}} \sqrt{\frac{2\sin^2\theta}{2\sin^2\theta+\rho\sigma_\xi^2 |s-\hat s|^2}}\\&\times\exp\left(
\frac{-\rho\mu_\xi^2|s-\hat s|^2}{4\sin^2\theta+2\rho\sigma_\xi^2 |s-\hat s|^2}
\right) d\theta.
\end{aligned}
\end{equation}
Let us define $\theta=\frac{\pi}{4}\varpi+\frac{\pi}{4}$, then the (\ref{epb1}) can be recast as
\begin{equation}\label{epb2}
\begin{aligned}
\bar P_b
=&\frac{1}{4}\int_{-1}^1 \sqrt{\frac{2\sin^2\left(\frac{\pi}{4}\varpi+\frac{\pi}{4}\right)}{2\sin^2\left(\frac{\pi}{4}\varpi+\frac{\pi}{4}\right)+\rho\sigma_\xi^2 |s-\hat s|^2}}\\&\times\exp\left(
\frac{-\rho\mu_\xi^2|s-\hat s|^2}{4\sin^2\left(\frac{\pi}{4}\varpi+\frac{\pi}{4}\right)+2\rho\sigma_\xi^2 |s-\hat s|^2}
\right) d\varpi.
\end{aligned}
\end{equation}
Let us define
$\varpi=\cos\left(\frac{2q-1}{2Q}\pi\right)$, then $\bar P_b$ can be further evaluated as
\begin{small}
\begin{equation}
\begin{aligned}
&\bar P_b
=\frac{\pi}{4Q}\sum_{q=1}^Q \sqrt{\frac{2\sin^2\left(\frac{\pi}{4}\cos\left(\frac{2q-1}{2Q}\pi\right)+\frac{\pi}{4}\right)}{2\sin^2\left(\frac{\pi}{4}\cos\left(\frac{2q-1}{2Q}\pi\right)\!+\!\frac{\pi}{4}\right)\!+\!\rho\sigma_\xi^2 |s-\hat s|^2}}\\&\times\exp\!\left(
\frac{-\rho\mu_\xi^2|s-\hat s|^2}{4\sin^2\left(\frac{\pi}{4}\cos\left(\frac{2q-1}{2Q}\pi\right)\!+\!\frac{\pi}{4}\right)\!+\!2\rho\sigma_\xi^2 |s\!-\!\hat s|^2}
\right) \!+\!R_Q,
\end{aligned}
\end{equation}
\end{small}
where $R_Q$ denotes the error term which can be neglected as the value of $Q$ is large.
\subsubsection{ Erroneous transmit antenna detection  $\hat n_t \neq n_t$}
According to (\ref{cpep2}), the UPEP can be written as
\begin{small}
\begin{equation}\label{cpep3n1}
\begin{aligned}
\bar P_b &= \int_0^\infty f(x)Q\left(\sqrt{\frac{\rho\left|G_{ n_t} s-G_{\hat n_t}\hat s\right|^2}{2}}\right)dx\\
&= \frac{1}{\pi}\int_0^\infty \int_0^{\frac{\pi}{2}} f(x)\exp\left(-\frac{\rho \left|G_{ n_t} s-G_{\hat n_t}\hat s\right|^2}{4\sin^2\theta}\right)d\theta dx\\
&= \frac{1}{\pi}\int_0^{\frac{\pi}{2}}\int_0^\infty f(x)\exp\left(-\frac{\rho \left|G_{ n_t} s-G_{\hat n_t}\hat s\right|^2}{4\sin^2\theta}\right)dxd\theta\\
&= \frac{1}{\pi}\int_0^{\frac{\pi}{2}}{\rm MGF}_\Gamma \left(\frac{-\rho }{4\sin^2\theta}\right) d\theta,
\end{aligned}
\end{equation}
\end{small}
where $f(x)$ denotes the PDF with respect the channel part in $\left|G_{ n_t} s-G_{\hat n_t}\hat s\right|^2$.

Although the integral variable in (\ref{cpep3n1}) contains only the channel component, the information in SM consists of two parts, the spatial domain and the symbol domain. Unfortunately, these two parts are coupled with each other in (\ref{cpep3n1}), which is challenging to resolve.
Nevertheless, in the following, we separate the information into two parts, real and imaginary, to address (\ref{cpep3n1}). Before that, we define
\begin{equation}\label{cpep3n2}
\begin{aligned}
\Gamma &= \left|G_{n_t}s-G_{\hat n_t}\hat s\right|^2\\
&=|\sum\nolimits_{l=1}^{L}\alpha_{l,n_t}\beta_ls-\sum\nolimits_{l=1}^L\alpha_{l,\hat n_t}\beta_{l}e^{-j(\theta_{l,\hat n_t}-\theta_{l,n_t})}\hat s|^2\\
&=|\sum\nolimits_{l=1}^{L}\beta_l\left(\alpha_{l,n_t}s-\alpha_{l,\hat n_t}e^{-j\phi_l}\hat s\right)|^2,
\end{aligned}
\end{equation}
where $\phi_l = \theta_{l,\hat n_t}-\theta_{l,n_t}$.
It can be observed that $s$, $e^{-j\phi_l}$, and $\hat s$ in (\ref{cpep3n2}) are complex values.
In this respect, we express (\ref{cpep3n2}) in terms of real and imaginary parts as
\begin{small}
\begin{equation}
\begin{aligned}
\Gamma
=&|\sum\nolimits_{l=1}^{L}\beta_l
\left(\alpha_{l,n_t}s_\Re-\alpha_{l,\hat n_t}(\cos\phi_l\hat s_\Re+\sin\phi_l\hat s_\Im)\right)
\\&
+j\sum\nolimits_{l=1}^{L}\beta_l\left(\alpha_{l,n_t}s_\Im-\alpha_{l,\hat n_t}(\cos\phi_l\hat s_\Im-\sin\phi_l\hat s_\Re)\right)|^2.
\end{aligned}
\end{equation}
\end{small}
To facilitate the representation, we define the real part information $\gamma_\Re$  and imaginary part information $\gamma_\Im$ within the absolute value as
\begin{small}
\begin{subequations}\label{cpep3n3}
\begin{align}
\gamma_\Re &= \sum\nolimits_{l=1}^{L}\beta_l\left(\alpha_{l,n_t}s_\Re-\alpha_{l,\hat n_t}(\cos\phi_l\hat s_\Re+\sin\phi_l\hat s_\Im)\right),\\
\gamma_\Im &= \sum\nolimits_{l=1}^{L}\beta_l\left(\alpha_{l,n_t}s_\Im-\alpha_{l,\hat n_t}(\cos\phi_l\hat s_\Im-\sin\phi_l\hat s_\Re)\right).
\end{align}
\end{subequations}
\end{small}
Resort to central limit theorem (CLT), both $\gamma_\Re$ and $\gamma_\Im$ obey a real Gaussian distribution. As such, it is our primary goal to determine the expectation and variance of $\gamma_\Re$ and $\gamma_\Im$.

On the one hand, since each element of RIS is independent of each other, the expectation of (\ref{cpep3n3}a) and (\ref{cpep3n3}b) can be respectively expressed as
\begin{small}
\begin{equation}\label{cpep3n4}
\begin{aligned}
&E[\gamma_\Re] \!=\! \sum\nolimits_{l=1}^{L}E[\beta_l\left(\alpha_{l,n_t}s_\Re\!-\!\alpha_{l,\hat n_t}(\cos\phi_l\hat s_\Re+\sin\phi_l\hat s_\Im)\right)]\\
&= \sum\nolimits_{l=1}^{L}E[\beta_l]E[\alpha_{l,n_t}s_\Re-\alpha_{l,\hat n_t}(\cos\phi_l\hat s_\Re+\sin\phi_l\hat s_\Im)]\\
&=\sum\nolimits_{l=1}^{L}E[\beta_l]E[\alpha_{l,n_t}s_\Re]-E[\alpha_{l,\hat n_t}(\cos\phi_l\hat s_\Re+\sin\phi_l\hat s_\Im)]\\
&=\sum\nolimits_{l=1}^{L}E[\beta_l]E[\alpha_{l,n_t}]s_\Re-E[\alpha_{l,\hat n_t}]E[\cos\phi_l\hat s_\Re\!+\!\sin\phi_l\hat s_\Im]\\
&=\sum\nolimits_{l=1}^{L}E[\beta_l]E[\alpha_{l,n_t}]s_\Re-E[\alpha_{l,\hat n_t}]\\
&\times\left(E[\cos\phi_l]\hat s_\Re+E[\sin\phi_l]\hat s_\Im\right),
\end{aligned}
\end{equation}
\begin{equation}\label{cpep3n5}
\begin{aligned}
&E[\gamma_\Im] \!=\! \sum\nolimits_{l=1}^{L}E[\beta_l\left(\alpha_{l,n_t}s_\Im\!-\!\alpha_{l,\hat n_t}(\cos\phi_l\hat s_\Im-\sin\phi_l\hat s_\Re)\right)]\\
&=\sum\nolimits_{l=1}^{L}E[\beta_l]E[\alpha_{l,n_t}s_\Im-\alpha_{l,\hat n_t}(\cos\phi_l\hat s_\Im-\sin\phi_l\hat s_\Re)]\\
&=\sum\nolimits_{l=1}^{L}E[\beta_l]E[\alpha_{l,n_t}s_\Im]-E[\alpha_{l,\hat n_t}(\cos\phi_l\hat s_\Im-\sin\phi_l\hat s_\Re)]\\
&=\sum\nolimits_{l=1}^{L}E[\beta_l]E[\alpha_{l,n_t}]s_\Im\!-\!E[\alpha_{l,\hat n_t}]E[\cos\phi_l\hat s_\Im-\sin\phi_l\hat s_\Re]\\
&=\sum\nolimits_{l=1}^{L}E[\beta_l]E[\alpha_{l,n_t}]s_\Im-E[\alpha_{l,\hat n_t}]\\&\times\left(E[\cos\phi_l]\hat s_\Im-E[\sin\phi_l]\hat s_\Re\right).
\end{aligned}
\end{equation}
\end{small}%
It can be observed that $\beta_l$, $\alpha_{l,n_t}$, and $\phi_l$ are the three main constituent components of (\ref{cpep3n4}) and (\ref{cpep3n5}), where $\beta_l$ and $\alpha_{l,n_t}$ denote that the magnitudes of $h_l$ and $g_{l,n_t}$ obeying the Rayleigh distribution. For this reason, we focus on finding the distribution of $\phi_l$ via the {\bf Lemma 1}.
\begin{lemma}
The PDF of variable $\phi_l$ can be expressed as
\begin{small}
\begin{equation}
f(\phi_l)=\left\{
\begin{aligned}
 & (\phi_l+2\pi)/(4\pi^2),  \ \ \  \phi_l\in[-2\pi,0]      \\
 & (-\phi_l+2\pi)/(4\pi^2),  \ \ \  \phi_l\in[0,2\pi]
\end{aligned}
\right.
\end{equation}
\end{small}
\end{lemma}
\emph{ Proof:} Please see the detail proof in \cite{zhucsi2022}.

\begin{equation}\small\label{phic1}
\begin{aligned}
E[\cos \phi_l]\!=\!&\int_{-2\pi}^0\!\frac{\phi_l\!+\!2\pi}{4\pi^2}\cos \phi_ld\phi_l\!+\!\int_0^{2\pi}\frac{-\phi_l\!+\!2\pi}{4\pi^2}\!\cos \phi_l d\phi_l\!=\!0,
\end{aligned}
\end{equation}
\begin{equation}\small\label{phis1}
\begin{aligned}
E[\sin \phi_l]\!=\!&\int_{-2\pi}^0\frac{\phi_l\!+\!2\pi}{4\pi^2}\sin \phi_ld\phi_l\!+\!\int_0^{2\pi}\frac{-\phi_l\!+\!2\pi}{4\pi^2}\sin \phi_l d\phi_l\!=\!0.
\end{aligned}
\end{equation}
Substituting (\ref{phic1}) and (\ref{phis1}) into (\ref{cpep3n4}) and (\ref{cpep3n5}), the mean values of $\gamma_\Re$ and $\gamma_\Im$ can be simplified to
\begin{equation}\label{phis2}
\begin{aligned}
E[\gamma_\Re]&=\sum\nolimits_{l=1}^{L}E[\beta_l\alpha_{l,n_t}]s_\Re={\pi Ls_\Re}/{4},\\
E[\gamma_\Im]&= \sum\nolimits_{l=1}^{L}E[\beta_l\alpha_{l,n_t}]s_\Im={\pi Ls_\Im}/{4}.
\end{aligned}
\end{equation}

On the other hand, the variance of (\ref{cpep3n3}a) and (\ref{cpep3n3}b) can be respectively expressed as
\begin{equation}\label{phixs3}
\begin{aligned}
Var[\gamma_\Re]&=E[\gamma_\Re^2]-E^2[\gamma_\Re],
Var[\gamma_\Im]=E[\gamma_\Im^2]-E^2[\gamma_\Im].
\end{aligned}
\end{equation}
The second moment of $\gamma_\Re$ can be evaluated as
\begin{small}
\begin{equation}\label{ere}
\begin{aligned}
&E[\gamma_\Re^2]=\sum_{l=1}^{L}E[\beta_l^2\left(\alpha_{l,n_t}s_\Re\!-\!\alpha_{l,\hat n_t}(\cos\phi_l\hat s_\Re+\sin\phi_l\hat s_\Im)\right)^2]\\
&=\sum_{l=1}^{L}E\left[\beta_l^2(\alpha_{l,n_t}^2|s_\Re|^2+\alpha_{l,\hat n_t}^2(\cos\phi_l\hat s_\Re+\sin\phi_l\hat s_\Im)^2)\right]\\&-2E[\alpha_{l,n_t}s_\Re\alpha_{l,\hat n_t}(\cos\phi_l\hat s_\Re+\sin\phi_l\hat s_\Im)].
\end{aligned}
\end{equation}
\end{small}
Based on (\ref{phic1}) and (\ref{phis1}), the (\ref{ere}) can be simplified to
\begin{equation}\label{ere1}
\begin{aligned}
&E[\gamma_\Re^2]
\!=\!\sum_{l=1}^{L}\!E[\beta_l^2\left(\alpha_{l,n_t}^2|s_\Re|^2\!+\!\alpha_{l,\hat n_t}^2(\cos\phi_l\hat s_\Re\!+\!\sin\phi_l\hat s_\Im)^2\right)]\\
&=\sum_{l=1}^{L}E\left[\beta_l^2(\alpha_{l,n_t}^2|s_\Re|^2\!+\!\alpha_{l,\hat n_t}^2(\cos^2\phi_l|\hat s_\Re|^2+\sin^2\phi_l|\hat s_\Im|^2\right.\\&\left.+2\cos\phi_l\hat s_\Re\sin\phi_l\hat s_\Im))\right].
\end{aligned}
\end{equation}
By means of (\ref{phic1}) and (\ref{phis1}), we can further obtain
\begin{small}
\begin{equation}
\begin{aligned}
&E[\gamma_\Re^2]
\!=\!\sum_{l=1}^{L}E\![\beta_l^2(\alpha_{l,n_t}^2|s_\Re|^2\!\\&+\!\alpha_{l,\hat n_t}^2(\cos^2\phi_l |\hat s_\Re|^2\!+\!\sin^2\phi_l|\hat s_\Im|^2))]\\
&=\sum\nolimits_{l=1}^{L}E[\beta_l^2(\alpha_{l,n_t}^2|s_\Re|^2+\alpha_{l,\hat n_t}^2\left((1+\cos2\phi_l)|\hat s_\Re|^2/{2}\right.\\&\left.+(1-\cos2\phi_l)|\hat s_\Im|^2/{2}\right))].
\end{aligned}
\end{equation}
\end{small}
After some calculations, we have
\begin{equation}\label{ere2}
\begin{aligned}
E[\gamma_\Re^2]
&=\sum\nolimits_{l=1}^{L}E[\beta_l^2(\alpha_{l,n_t}^2|s_\Re|^2+\alpha_{l,\hat n_t}^2\frac{|\hat s|^2}{2})].
\end{aligned}
\end{equation}
Since $\beta_l$, $\alpha_{l,n_t}$, and $\alpha_{l,\hat n_t}$ follow Rayleigh distribution, we can get
$
E[\beta_l^2]=E^2[\beta_l]+Var[\beta_l]=\frac{\pi}{4}+\frac{4-\pi}{4}=1$ and
$E[\alpha_{l,n_t}^2]=E[\alpha_{l,\hat n_t}^2]=1.$
By applying the CLT, we have
$
E[\gamma_\Re^2]=
|s_\Re|^2L+\frac{|\hat s|^2L}{2}.
$
Here, we can calculate the variance of $\gamma_\Re$ as
\begin{equation}
Var[\gamma_\Re]
={(16-\pi^2)|s_\Re|^2L}/{16}+{|\hat s|^2L}/{2}.
\end{equation}
Similarly, we have
\begin{equation}
Var[\gamma_\Im]
={(16-\pi^2)|s_\Im|^2L}/{16}+{|\hat s|^2L}/{2}.
\end{equation}
Since $E[\gamma_\Re]$ and $E[\gamma_\Im]$ are obtained, then we derive the $E[\gamma_\Re\gamma_\Im]$ as
\begin{small}
\begin{equation}\label{covd1}
\begin{aligned}
&E[\gamma_\Re\gamma_\Im]
\!=\! \sum_{l=1}^{L}E\!\left[\beta_l^2\left(\alpha_{l,n_t}s_\Re\!-\!\alpha_{l,\hat n_t}(\cos\phi_l\hat s_\Re\!+\!\sin\phi_l\hat s_\Im)\right)\right.\\&\left.\times\left(\alpha_{l,n_t}s_\Im-\alpha_{l,\hat n_t}(\cos\phi_l\hat s_\Im-\sin\phi_l\hat s_\Re)\right) \right]\\
=& \sum\nolimits_{l=1}^{L}E\left[\left(\alpha_{l,n_t}s_\Re-\alpha_{l,\hat n_t}(\cos\phi_l\hat s_\Re+\sin\phi_l\hat s_\Im)\right)\right.\\&\times\left.\left(\alpha_{l,n_t}s_\Im-\alpha_{l,\hat n_t}(\cos\phi_l\hat s_\Im-\sin\phi_l\hat s_\Re)\right) \right]\\
=&\sum\nolimits_{l=1}^{L}E\left[s_\Re s_\Im +\cos^2\phi_l\hat s_\Re\hat s_\Im-\sin^2\phi_l\hat s_\Re\hat s_\Im\right.\\&\left.
-\cos\phi_l\hat s_\Re\sin\phi_l\hat s_\Re+\sin\phi_l\hat s_\Im\cos\phi_l\hat s_\Im\right]\\
=& \sum\nolimits_{l=1}^{L}E\left[s_\Re s_\Im +\cos^2\phi_l\hat s_\Re\hat s_\Im-\sin^2\phi_l\hat s_\Re\hat s_\Im
\right]\\
=& \sum\nolimits_{l=1}^{L}E\left[s_\Re s_\Im
\right]= s_\Re s_\Im L.
\end{aligned}
\end{equation}
\end{small}%
Then, the covariance of $\gamma_\Re$ and $\gamma_\Im$ can be expressed as
\begin{equation}
Cov[\gamma_\Re\gamma_\Im]=s_\Re s_\Im L-\frac{\pi^2s_\Re s_\Im L}{16}=\frac{(16-\pi^2)s_\Re s_\Im L}{16}.
\end{equation}
Let us define $\mathbf{\Gamma}=\mathbf{z}^T\mathbf{Az}$, where $\mathbf{z} = [\gamma_\Re \ \ \gamma_\Im]^T$ and $\mathbf{A}=\mathbf{I}_2$.
The mean vector and covariance matrix of $\mathbf{z}$ are described as
\begin{small}
\begin{equation}
\boldsymbol{\mu}=
\begin{bmatrix}
\frac{L\pi}{4}s_\Re & \frac{L\pi}{4}s_\Im
\end{bmatrix},
\end{equation}
\begin{equation}
\mathbf{V}=
\begin{bmatrix}
Var[\gamma_\Re] & Cov[\gamma_\Re\gamma_\Im]\\
Cov[\gamma_\Im\gamma_\Re] & Var[\gamma_\Im]
\end{bmatrix}.
\end{equation}
\end{small}%
According to \cite{basar2020rec}, the MGF of $\Gamma$ can be calculated by
\begin{small}
\begin{equation}\label{nmgf}
\begin{aligned}
&{\rm MGF}_\Gamma(x)=\left({\rm det}(\mathbf{I}-2x\mathbf{AV})\right)^{-\frac{1}{2}}\\&\times\exp(
-\frac{1}{2}\boldsymbol{\mu}^H\left(\mathbf{I}-(\mathbf{I}-2x\mathbf{AV})^{-1}\right)\mathbf{V}^{-1}\boldsymbol{\mu}
).
\end{aligned}
\end{equation}
\end{small}%
In this stage, we consider (\ref{cpep3n1}) and (\ref{nmgf}).
Hence, the UPEP can be given by
\begin{small}
\begin{equation}
\begin{aligned}
&\bar P_b
=\frac{1}{\pi}\int_0^{\frac{\pi}{2}}({\rm det}(\mathbf{I}+\frac{\rho}{2\sin^2\theta}\mathbf{AV}))^{-\frac{1}{2}}\\&\times\exp(
-\frac{1}{2}\boldsymbol{\mu}^H(\mathbf{I}-(\mathbf{I}+\frac{\rho}{2\sin^2\theta}\mathbf{AV})^{-1})\mathbf{V}^{-1}\boldsymbol{\mu}
) d\theta.
\end{aligned}
\end{equation}
\end{small}%
To obtain closed-form expression of UPEP, we deal with them with the GCQ method.
Let $\theta=\frac{\pi}{4}\varpi+\frac{\pi}{4}$, after some calculations, the UPEP in this case can be written as
\begin{small}
\begin{equation}
\begin{aligned}
&\bar P_b
=\frac{\pi}{4Q}\sum\nolimits_{q=1}^Q ({\rm det}(\mathbf{I}+\frac{\rho}{2\sin^2(\frac{\pi}{4}\varpi+\frac{\pi}{4})}\mathbf{AV}))^{-\frac{1}{2}}\\
&\times\exp(
-\frac{1}{2}\boldsymbol{\mu}^H(\mathbf{I}-(\mathbf{I}+\frac{\rho}{2\sin^2(\frac{\pi}{4}\varpi+\frac{\pi}{4})}\mathbf{AV})^{-1})\mathbf{V}^{-1}\boldsymbol{\mu}
)+R_Q,
\end{aligned}
\end{equation}
\end{small}%
where $\varpi=\cos\left(\frac{2q-1}{2Q}\pi\right)$ and $R_Q$ denotes the error term that can be ignored, when the $Q$ value is relatively large.
\subsection{ABEP}
In general, there is no precise ABEP expression for RIS-SM systems for arbitrary spatial and symbol domain modulation orders.
Consequently, the union upper bound of ABEP on the RIS-SM system can be characterized as
\begin{small}
\begin{equation}\label{abep}
ABEP \leq \sum_{n_t = 1}^{N_t}\sum_{m=1}^M\sum_{\hat n_t = 1}^{N_t}\sum_{\hat m=1}^M\frac{\bar P_b N([n_t, m]\to [\hat n_t, \hat m])}{N_tM\log_2(N_tM)},
\end{equation}
\end{small}%
where $N([n_t, m]\to [\hat n_t, \hat m])$ denotes the Hamming distance between the transmitted signal indices from $n_t$ to $\hat n_t$ and to the detected signal from $s$ to $\hat s$.
It is worth noting that the equation sign holds if and only if $N_t=2$ and $M=1$.

\section{Simulation and Analytical Results}
In this section, we assess the ABEP of the RIS-SM scheme for downlink transmission over the Rayleigh channel.

\begin{figure}[t]
  \centering
  \includegraphics[width=5.5cm]{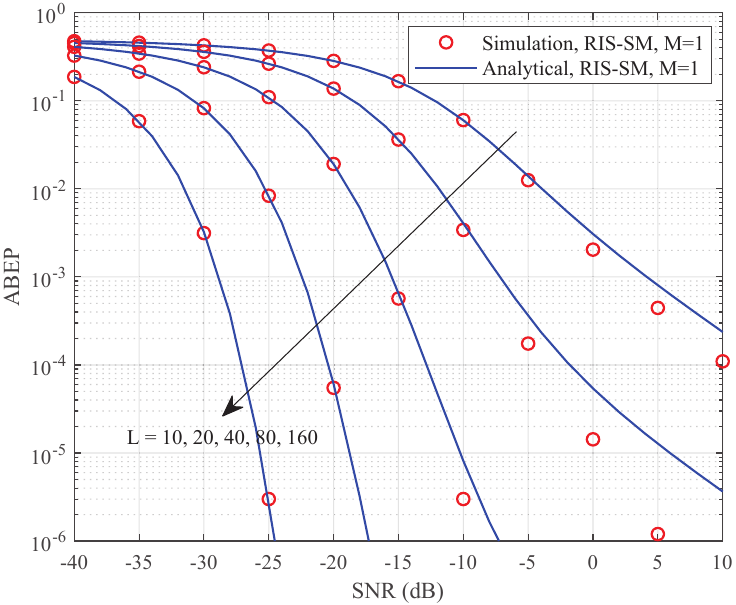}\\
  \caption{Verification of the CLT-based analytical ABEP results.}\label{CLT}
\end{figure}

In Fig. \ref{CLT}, we validate the theoretical ABEP results obtained using the CLT approach against simulation results, with $N_t=2$ and $M=1$. The convergence between theoretical and simulation results is evident when the number of reflective elements in the RIS is 80 and 160. However, the gap between the two widens when the number of RIS elements is 10, 20, and 40. This discrepancy arises because the composite channel distribution more accurately approximates the Gaussian distribution when the number of reflecting elements $L$ is equal to or exceeds 80.

\begin{figure}[t]
  \centering
  \includegraphics[width=5.5cm]{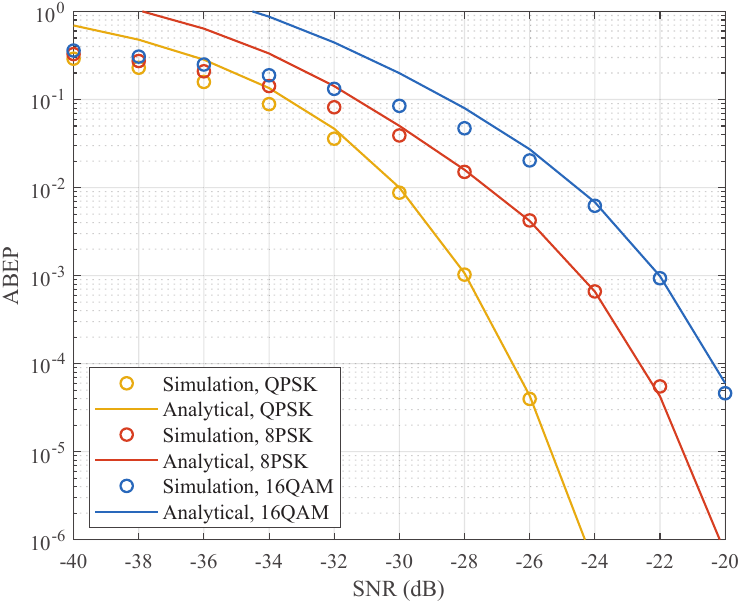}\\
  \caption{ABEP performance under different modulation orders.}\label{PSKQAM}
\end{figure}

In Fig. \ref{PSKQAM}, we illustrate the ABEP performance versus SNR results for the RIS-SM system under different modulation orders. Here, the number of transmit antennas and the number of reflecting elements in the RIS are set to 2 and 100, respectively.
Observing Fig. \ref{PSKQAM}, it is evident that in the low SNR region, simulation results and analytical curves exhibit some disparity. However, with increasing SNR, the two results gradually converge and become consistent. Additionally, we note that the performance of the RIS-SM system degrades as the modulation order in the symbol domain increases. This deterioration is attributed to the normalization of the energy of the transmitted symbol $s$. With higher modulation orders, the Euclidean distance between adjacent constellation points decreases, impacting system performance.


\begin{figure}[t]
  \centering
  \includegraphics[width=5.5cm]{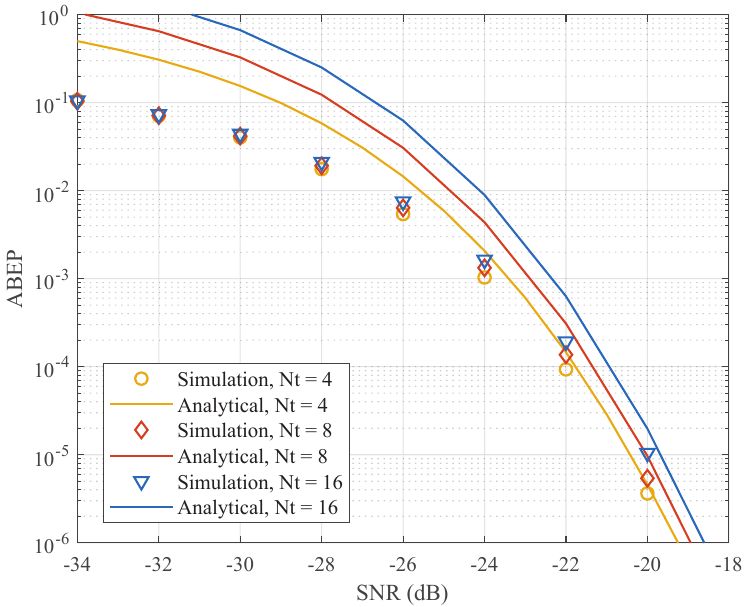}\\
  \caption{ABEP performance with different transmit antennas.}\label{NtBER}
\end{figure}

In Fig. \ref{NtBER}, we present the ABEP performance curves alongside Monte Carlo simulation results for the proposed RIS-SM system. The modulation order and the number of reflecting units in the RIS are specified as 2 and 100, respectively.
The results show that as the SNR increases, the simulation results converge with the analytical results.
Additionally, we observe that the ABEP performance of the RIS-SM system exhibits minimal degradation even with an increase in the number of transmit antennas from 4 to 16.

\begin{figure}[t]
  \centering
  \includegraphics[width=5.5cm]{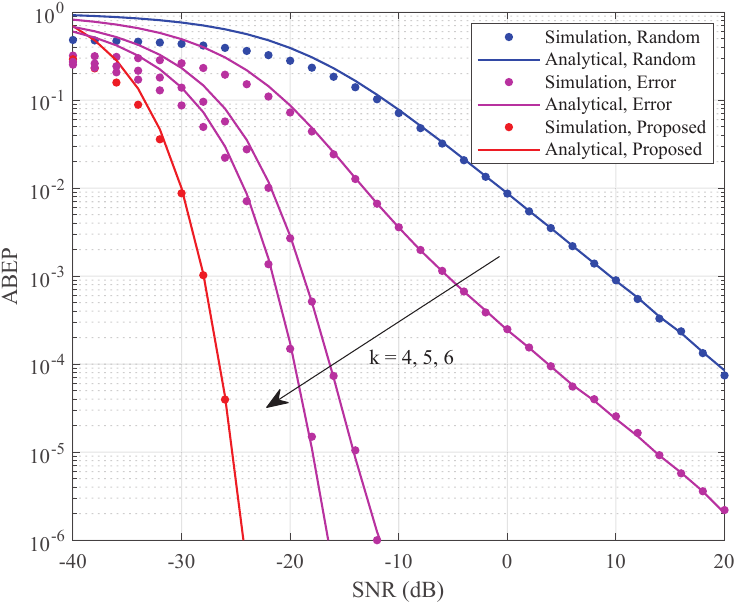}\\
  \caption{The impact of RIS phase accuracy on ABEP.}\label{error}
\end{figure}

In Fig. \ref{error}, we investigate the ABEP performance of the RIS-SM scheme under phase adjustment accuracy on RIS. The parameters are set as $M=2, N_t=2$, and $L=100$. Specifically, the blue curve in Fig. \ref{error} represents randomly generated reflecting shifts, while the purple curves depict cases with adjustment errors in the reflecting phase shift.
For simplicity, we assume that the interval of the phase shift adjustment error is uniformly distributed between $\left[-\pi/k, \pi/k\right]$. We observe that in the medium and high SNR regions, the analytical and simulation results are highly matched.

\section{Conclusion}

This paper introduced a novel RIS-assisted SM downlink communication system and conducted a thorough analysis of its reliability. Leveraging the GCQ method and CLT, we derived a closed-form expression for the union tight upper bound of ABEP. The validity of the derived results is extensively confirmed through Monte Carlo simulations. Additionally, we assessed the applicability conditions of GCQ and CLT. Furthermore, we explored the impact of the number of antennas, modulation order, and phase adjustment accuracy  of the RIS on the performance of the RIS-SM system.

\section{Acknowledgment}
Q. Wu's work is supported by National Key R\&D Program of China (2022YFB2903500), NSFC 62371289 and NSFC 62331022.
W. Chen's work is supported by National key project 2020YFB1807700, NSFC
62071296, Shanghai 22JC1404000, 20JC1416502, and PKX2021-D02.

\end{document}